\documentstyle[aps,epsfig,twocolumn]{revtex}
\begin{document}
\draft
\flushbottom
\twocolumn[
\hsize\textwidth\columnwidth\hsize\csname @twocolumnfalse\endcsname

\title{Generalization of Child-Langmuir Law for Non-Zero Injection
Velocities in a Planar Diode}

\author{R.~R.~Puri, Debabrata Biswas and Raghwendra Kumar}
\address{
Theoretical Physics Division \\
Bhabha Atomic Research Centre \\
Mumbai 400 085, INDIA}
\date{\today}
\maketitle


\begin{abstract}
The Child-Langmuir law relates the voltage applied across a planar
diode to the saturation value $J_{{\rm CL}}$ of current density
that can be transmitted through it in case the injection velocity of
electrons into the diode is zero. The Child-Langmuir current density
$J_{{\rm CL}}$ is, at the same time, (i) the maximum current density
that can be transmitted through a planar diode, (ii) the current
density below which the flow is steady and unidirectional in the long
time limit and (iii) the average transmitted current density for
{\em any} value of injected current density above
$J_{{\rm CL}}$. Existing generalizations of Child-Langmuir law
to non-zero velocities of injection are based on the characteristics
(i) and (ii) of $J_{{\rm CL}}$. This paper generalizes the law to
non-zero velocities of injection based on the characteristic (iii) by
deriving an analytical expression for the saturation value of current
density. The analytical expression for the saturation current density
is found to be well supported by numerical computations. A reason
behind preferring the saturation property of the Child-Langmuir current
density as the basis for its generalization is the importance of that
property in numerical simulations of high current diode devices.
\vskip 0.1 in
\end{abstract}


\date{\today}
]
\narrowtext
\tightenlines

\newcommand{\be}{\begin{equa}}
\newcommand{\ee}{\end{equation}}
\newcommand{\bea}{\begin{eqnarray}}
\newcommand{\eea}{\end{eqnarray}}
\newcommand{\Tbar}{{\overline{T}}}
\newcommand{\En}{{\cal E}}
\newcommand{\Lop}{{\cal L}}
\newcommand{\DB}[1]{\marginpar {\footnotesize {\em #1 }}}
\newcommand{\DBT}[1]{\marginpar {\footnotesize \tighten {\em #1 }}}
\newcommand{\q}{\vec{q}}
\newcommand{\kt}{\tilde{k}}
\newcommand{\Lopn}{\tilde{\Lop}}

\pagebreak
\section{Introduction}

The celebrated Child-Langmuir law is widely used in many areas of
physics where charged particle transport across a potential difference
is involved~\cite{ender}. It gives the maximum current density
$J_{\rm CL}$ that can be transported from an infinite planar cathode
at zero potential to an infinite planar anode parallel to the cathode
at a distance $L$ and potential $V$ when the initial velocity $v_0$ of
electrons at the cathode is zero. This maximum value is a result of
the space charge effect and, in the non-relativistic limit, is given
by
\bea
J_{\rm CL} = {V^{3/2} \over L^2} {1 \over 9 \pi}
\left({2e\over m}\right)^{1/2}.
\label{eq:CL1}
\eea
\noindent
Here, $e$ is the magnitude of the charge on an electron and
$m$ is its rest mass. For convenience, the terms ``current'' and
``current density'' are used interchangeably. They refer to the
current density denoted by $J$.

There have been several attempts to generalize the law (\ref{eq:CL1})
to other situations such as to higher dimensions or to non-zero
velocities of injection \cite{CL2D}-\cite{liu}.
The characteristics of $J_{{\rm CL}}$ on
which such generalizations are based are one or the other of the
following: (i) $J_{{\rm CL}}$ is the maximum current that can be
transmitted across one-dimensional planar diode with zero velocity of
injection, and (ii) the flow of current in the long time limit
is steady and unidirectional
if the injected current is less than $J_{{\rm CL}}$ but oscillatory in
time above it. The generalization based on characteristic (i) to
non-zero injection velocity, $v_0$, of electrons in to a planar
diode shows that the space charge limiting current 
\bea
J_{\rm SCL}&=&J_{\rm CL} \left[\left({mv_0^2\over 2eV}\right)^{1/2} +
\left(1 + {mv_0^2\over 2eV} \right)^{1/2} \right]^3
\label{eq:scl}
\eea
is the maximum current that can be transmitted across the diode and
is identified as the generalized Child-Langmuir current~\cite{akimov}.
The generalization based on characteristic (ii), on the other hand,
shows that the current in the diode is unidirectional till it attains
the value $J_{\rm BF}$ and starts reflecting
thereafter with 
\bea
J_{\rm BF}&=&J_{\rm CL} \left[\left({mv_0^2\over 2eV}\right)^{3/4} +
\left(1 + {mv_0^2\over 2eV} \right)^{3/4} \right]^2
\label{eq:bf}
\eea
This current, $J_{BF}$, marking the bifurcation of the state from
completely to partially transmitting is proposed in Ref.~\cite{liu}
as a generalized Child-Langmuir current.

A third characteristic that is commonly ignored in such attempts is
the fact that $J_{{\rm CL}}$ is the saturation current that is
transmitted for zero velocity of injection for {\it any}
injected current higher than $J_{{\rm CL}}$. It is the existence of
saturation current in a planar diode that we wish to explore in this
communication for non-zero velocities of injection and call the
relation between such a saturation current, the injection velocity,
the diode spacing and voltage as the generalized Child-Langmuir law.

The reason for emphasising the saturation property is that it helps
in simulations involving high current devices. For, the existence of a
saturation current implies that one need not bother about the material
dependent cathode characteristics (the Richardson-Dushman law in case of
thermionic emission or the Fowler-Nordheim law for field emission) so
long as the current emitted from the cathode surface is more than what
gives rise to the saturation value of the transmitted current. This
condition simplifies considerably the task of numerical
simulation.

Though the injection velocity of electrons in many devices can be
taken as zero, in several others non-zero injection velocities do
arise (see~\cite{ender} and references therein). The magnitude of the
velocity can be related to physically measurable quantities such as
the temperature in case of thermionic emission or the frequency of the
incident field in  case of photoemission. Since, as in the case of
$v_0=0$, existence of a saturation current for $v_0\ne 0$ may enable
one to ignore the cathode characteristics for appropriate injected
currents, it is important to know for what values of injected current
and to what value, if any, does the transmitted current saturate if
$v_0\ne 0$. The main result of this paper is that as the current
injected in to a planar diode is increased beyond its space charge
limiting value, the current transmitted through it saturates
{\it asymptotically} to the value 
\bea
J_{\rm GCL}&=& J_{\rm CL} \left(1+{mv^2_0\over 2eV}\right)^{3/2},
\label{eq:gcl}
\eea
which we call the generalized Child-Langmuir current. The equation
above is the desired generalization to $v_0\ne 0$ of the
Child-Langmuir relation (\ref{eq:CL1}) for $v_0=0$.
Note that $J_{\rm GCL}\le J_{\rm BF}\le J_{\rm SCL}$ and that the
equalities hold when $v_0=0$.

The relation (\ref{eq:gcl}) is derived in the framework of a
phenomenological theory, called the classical theory.
The validity of that theory is assessed by comparison with
numerical solution of the equations.

The paper is organised as follows. In Sec.2 we recall the main results
of the steady state theory leading to the expression for the space
charge limiting current. Some details of derivation of those results
are outlined in the Appendix A. The classical theory for currents
exceeding the space charge limiting current is presented in Sec.3.
The results of numerical computation are presented in Sec.4.

\section{Space Charge Limiting Current}
\label{sec:limiting}

Consider a diode consisting of two infinite parallel plates placed at
$z=0$ and $z=L$ where $z$ is the direction orthogonal to the plates.
Let an electronic fluid of uniform and same cross-section as the
plates move along the z-direction. Its motion in the Eulerian
formalism is governed by the equations~\cite{birdsall}
\bea
&&{\partial^2\phi\over\partial z^2}=
-{\partial E\over\partial z}=-4\pi\rho(z,t),\nonumber\\
&&{\partial\rho\over\partial t}+{\partial J\over\partial z}=0,
\nonumber\\
&&{{\rm d}p\over{\rm d}t}=-eE(z,t),
\label{2.1}
\eea
where $\rho(z,t)$ and $J(z,t)$ are respectively the charge and current
densities of the electronic fluid, $p(z,t)$ is the momentum of the
electron whereas $\phi(z,t)$ and $E(z,t)$ are the potential and the
electric field at the position $z$ at time $t$.

We consider the case when a stream of electrons of constant charge
density $\rho_0$ enters the diode at $z=0$ with each electron in the
stream entering with the same velocity $v_0$ along the $z$-direction
so that
\bea
\rho(0,t)=\rho_0,\qquad J(0,t)\equiv\rho(0,t)v_0=J_0.
\label{2.5}
\eea
Furthermore, the plates at $z=0$ and $z=L$ are assumed to be
maintained at fixed potentials given by
\bea
\phi(0,t)=0,\qquad \phi(L,t)=V.
\label{2.6}
\eea
Because of the defining relation between the field and the potential,
the condition (\ref{2.6}) may be rewritten as
\bea
\int_{0}^{L}E(z,t){\rm d}z=V.
\label{2.7}
\eea
A complete description of the motion of the electronic fluid is
provided by solving equations (\ref{2.1}) under the
boundary condition (\ref{2.6}) or (\ref{2.7}).

A state of interest, called a steady state, is the one that the
system reaches asymptotically in time i.e. as $t\rightarrow\infty$.
The evaluation of space charge limiting current requires solving the
steady state form of the equations obtained by equating to zero the
partial derivatives with respect to time in the Eulerian formalism.
It is evident from (\ref{2.1}) that the current density in this case
is not only independent of time but is also the same at every plane.
If there is no reflection of electrons from anywhere in the diode then
the current density in the diode is everywhere the same as is pumped
in at its entrance so that, in the steady state,
$J(z,t)\equiv \rho(z,t)v(z,t)=J_0$. The first of the equations in
(\ref{2.1}) then assumes the form
\bea
{{\rm d}^2\phi\over{\rm d}z^2}=-{4\pi J_0\over v}.
\label{3.2}
\eea
The relation between $\phi$ and $v$ is provided by the steady state
form of the law of conservation of energy,
\bea
{1\over 2}mv^2-e\phi={1\over 2}mv^2_0\equiv\varepsilon
\label{3.3}
\eea
where we have introduced the parameter $\varepsilon$ to denote the
electronic energy and assumed that $\varepsilon$ is small enough for
the motion to be considered non-relativistic. On combining (\ref{3.2})
and (\ref{3.3}) we arrive at the equation
\bea
{{\rm d}^2\bar\phi\over{\rm d}\bar z^2}=
{4\alpha\over 9\sqrt{\bar\phi}}
\label{3.4}
\eea
where the scaled current $\alpha$ and the scaled potential
$\bar\phi$ are defined by
\bea
\alpha&=&9\pi e|J_0|L^2\varepsilon^{-3/2}\sqrt{{m\over 2}},
\label{alpha}\\
\bar\phi&=&1+{e\phi\over\varepsilon}.
\label{3.5}
\eea
In terms of the scaled potential, the boundary condition (\ref{2.6})
assumes the form
\bea
\bar\phi(z=0,t)=1,\qquad \bar\phi(z=L,t)=\bar V,
\label{3.5b}
\eea
where
\bea
\bar V=1+{eV\over\varepsilon}.
\label{3.5c}
\eea
Though the steady state may be determined by solving (\ref{3.4})
along with the boundary conditions (\ref{3.5b}), it turns out to be
more convenient to derive it instead in the Lagrangian formalism. We
will, however, return to Eq.(\ref{3.4}) in the next section which
concerns the situation involving returning currents.

Alternative to the Eulerian formalism which characterizes the motion
in terms of hydrodynamical variables like charge and current
densities as function of space and time, is the Lagrangian formalism
which is based on determining the trajectories of the electrons. An
electronic trajectory in the one-dimensional motion under
consideration is identified by the time $t_0$ at which it enters the
diode. In the non-relativistic limit ($p=mv$), the trajectory of
an electron entering the diode at $t=t_0$ is determined by the
solution of the Llewellyn's equation
\bea
{\partial^3 z(t,t_0)\over\partial t^3}=
-{e\over m}\left[{\partial E_0(t)\over\partial t}
+4\pi J_0(t)\right],
\label{2.9}
\eea
where $E_0(t)\equiv E(0,t)$ and $J_0(t)\equiv J(0,t)$ are the electric
field and the current density at the entrance ($z=0$) of the diode.
The steady state solution of this equation has been derived in
Ref.~\cite{akimov}. However, for the sake of completeness and clarity,
and in order to fill the gaps in the available derivations, we present
in Appendix A the solution of (\ref{2.9}) for the steady state
corresponding to unidirectional flow of current in terms of the
following dimensionless variables
\bea
\bar z&=&{z\over L},\qquad
\bar t={t\over L}\sqrt{2\varepsilon\over m},\qquad
\bar v=\sqrt{{m\over 2\varepsilon}}v={v\over v_0},\nonumber\\
\bar E&=&{eL E\over \varepsilon}=
-{{\rm d}\bar\phi\over{\rm d}\bar z}.
\label{2.13}
\eea
The injected current $J_0$ in the steady state is transmitted
completely to the anode where the electrons arrive with
velocity $v_1$ which, by virtue of the energy conservation relation
(\ref{3.3}), is given by
\bea
\bar v_1=\sqrt{\bar V}.
\label{3.8}
\eea
The main results of the steady state solution are:

\noindent
\begin{enumerate}

\item The steady state is determined by solving the cubic (\ref{3.11})
for the transit time $\bar T(1)$ of an electron across the diode.
The flow of current is unidirectional and independent of time if that
cubic admits physically acceptable solution.

\item The cubic admits two positive roots, $\bar T_2(1)$ and
$\bar T_3(1)$, given in $(\ref{3.16})$, if the value of the injected
current $\alpha\le\alpha_{{\rm SCL}}$ where
\bea
\alpha_{{\rm SCL}}=(1+\sqrt{\bar V})^3
\label{n1}
\eea
is the space charge limiting current. By invoking (\ref{alpha}),
it may be verified that the expression for the injected current
$J_0$ corresponding to $\alpha$ given above is the same as in
(\ref{eq:scl}) and that if $v_0=0$ then
\bea
J_{{\rm SCL}}=J_{{\rm CL}},\qquad v_0=0,
\label{n1n}
\eea
where $J_{{\rm CL}}$ is given by (\ref{eq:CL1}). In other words,
for zero velocity of injection, the space charge limiting current is
the same as the Child-Langmuir current.

\item For $0\le\alpha\le\alpha_{{\rm DR}}$, the root $\bar T_3(1)$
corresponds to negative velocities in certain region of the diode and
hence is unacceptable in the said range of the injected current.
However, both the roots, $\bar T_2(1)$ and $\bar T_3(1)$ are
acceptable if $\alpha_{{\rm DR}}\le\alpha\le\alpha_{{\rm SCL}}$ where
\bea
\alpha_{{\rm DR}}=(1+\bar V^{3/4})^2.
\label{n1p}
\eea
The suffix ${\rm DR}$ on $\alpha$ stands for "double root": it
is the value of the current above which steady state equation has
two admissible roots. The linear stability analysis shows that, in the
range mentioned above, $\bar T_3(1)$ leads to unstable steady state.
Hence, $\bar T_2(1)$ defines the acceptable stable steady state for
all $\alpha\le\alpha_{{\rm SCL}}$.

\item The flow of current ceases to be unidirectional and
time-independent if the injected current exceeds $\alpha_{{\rm SCL}}$.
\end{enumerate}

\noindent
The above-mentioned characteristics of the steady state solution are
exhibited in Fig.1 by the plot of the electric field $\bar E_0$ at the
cathode evaluated using (\ref{3.17b}) for $\bar V=1.2$. The point $A$
in the figure corresponds to the space charge limiting current
$\alpha_{{\rm SCL}}$ whereas the point $B$ represents
$\alpha_{{\rm DR}}$. The branches OA and AB in that figure correspond,
respectively, to $\bar T_2(1)$ and $\bar T_3(1)$ given by
(\ref{3.16}). The branch AB represents unstable steady state. The
uppermost branch of the curve for $\alpha>\alpha_{{\rm DR}}$ is a
result of the classical theory outlined in the next section. The value
of $\bar E_0$ at $\alpha_{{\rm DR}}$ where the unstable steady state
corresponding to $\bar T_3(1)$ merges with the classical theory, found
using (\ref{3.17b}), is given by
\bea
\bar E_{0{\rm DR}}={4\over 3}(1+\bar V^{3/4}).
\label{efdr}
\eea
We will see that, as it should be, this value is the same as that
obtained by the classical theory.

By transforming back to the unnormalized variables, the results
above may be applied to the case of $v_0=0$. It may be
verified that the expression for space charge limiting current density
$J_0$, found using (\ref{n1}) in this case is the well-known
Child-Langmuir current $J_{{\rm CL}}$ given in (\ref{eq:CL1}) and
that, correspondingly, $E_0=0$. We thus see that, contrary to the
standard treatments of the case of $v_0=0$, $E_0=0$ is
not an imposed condition but is a natural consequence of the governing
equations of motion. This fact has also been emphasised
in~\cite{akimov}.

Since the space charge limiting current $\alpha_{{\rm SCL}}$ is the
maximum current that can be transmitted across the diode, it is
proposed in~\cite{akimov} that $\alpha_{{\rm SCL}}$ may be identified
as the generalized Child-Langmuir current and (\ref{n1}) as the
generalized Child-Langmuir current-voltage relation. However, as we
have argued in the Introduction, an important characteristic of
Child-Langmuir current is that it is the current transmitted across
the diode for any injected current above $\alpha_{{\rm SCL}}$.
We, therefore, need to examine the behaviour of the transmitted
current for $\alpha>\alpha_{{\rm SCL}}$.

As shown in the Appendix A, the system does not admit a steady state
corresponding to unidirectional flow of current if
$\alpha>\alpha_{{\rm SCL}}$. Hence, the behaviour of the system for
$\alpha>\alpha_{{\rm SCL}}$ is determined by solving time-dependent
equations. Since solving time-dependent equations analytically is a
formidable task, those equations are solved numerically. However, a
qualitative understanding of the phenomenon is achieved by a simple
model which assumes that, though for $\alpha>\alpha_{{\rm SCL}}$ the
current may not be unidirectional, it nevertheless attains a steady
state with current flowing in two directions in some region in the
diode. This theory, called the classical theory, is outlined in the
next section. It is along the lines of the theory in
Ref.~\cite{birdsall} for the case of grounded plates.

\section{Asymptotically Saturating Current}

The analysis outlined in the Appendix A holds good for
$\alpha\leq \alpha_{{\rm SCL}}$ in which case the injected and the
transmitted currents are equal. The classical theory assumes that,
beyond $\alpha_{{\rm SCL}}$, the minima of the potential within the
diode assumes a value which is such that the velocity of an electron
there reduces to zero. The equations (\ref{3.3}) and (\ref{3.5}) show
that $v=0$ is attained when $\bar\phi=0$. Hence, the boundary
condition at the position $\bar z=\bar z_m$ of the potential minima is
\bea
\bar\phi={{\rm d}\bar\phi\over{\rm d}\bar z}=0,\qquad
(\bar z=\bar z_m).
\label{3.24}
\eea
After the electrons come to rest at $\bar z=\bar z_m$, it is assumed
that a fraction $f$ of them is transmitted while the remaining
fraction $1-f$ is reflected back towards the cathode. Thus, in the
region $0\le\bar z\le \bar z_m$, there are two currents: the injected
current $\alpha$ moving away from the cathode and $(1-f)\alpha$ moving
towards it. In the region $\bar z_m\le\bar z\le 1$, there is
unidirectional flow of current of magnitude $f\alpha$ towards the
anode. Note that the net current at every plane in the two regions is
$f\alpha$. Since the charge densities at a point due to different
currents at it add, the Poisson's equation (\ref{3.4}) in the two
regions assumes the forms
\bea
{{\rm d}^2\bar\phi_L\over{\rm d}\bar z^2}&=&{4\alpha(2-f)
\over 9\sqrt{\bar\phi_L}},\qquad ~~~~0\le\bar z\le \bar z_m,
\label{3.25_1}\\
{{\rm d}^2\bar\phi_R\over{\rm d}\bar z^2}&=&{4\alpha f
\over 9\sqrt{\bar\phi_R}},~~~~~~\qquad \bar z_m\le \bar z\le 1.
\label{3.25}
\eea
Both these equations are of the form ${\rm d}^2 \phi/{\rm d}z^2 =
C/\phi^{1/2}$ and can be recast as

\bea
{{\rm d}\over {\rm d}z}\left({{\rm d}\bar\phi\over{\rm d}\bar z}
\right)^2 = 4C {{\rm d}\sqrt{\bar\phi} \over{\rm d}\bar z}.
\label{3.26}
\eea
The solution of this equation, satisfying the boundary condition
(\ref{3.24}) reads,
\bea
{{\rm d}\bar\phi\over{\rm d}\bar z}=\pm 2\sqrt{C}\bar\phi^{1/4}.
\label{3.28}
\eea
Now, recall that in a one-dimensional electrostatic situation involving
charges of only one sign, there can be only one extrema. In the
problem at hand, the minima at $\bar z=\bar z_m$, therefore, is the
only one admissible so that the electrons are decelerated while moving
from the cathode to the position of minima at $\bar z=\bar z_m$ and
are accelerated thereafter. Hence, the minus sign in the solution
above holds in the region $0\le\bar z<\bar z_m$ whereas the plus sign
is to be retained in the region $\bar z_m<\bar z<1$. On integrating
(\ref{3.28}) in the two regions with appropriate values of $C$ along
with the boundary conditions at $\bar z=0,1$ given in (\ref{3.5b}), it
follows that
\bea
\bar\phi_L^{3/4} & = & 1-\bar z\sqrt{\alpha (2-f)},
~~~~~~~~0 \leq\bar z \leq \bar z_m,
\label{3.30}\\
\bar\phi_R^{3/4} & = &(\bar z-1)\sqrt{\alpha f}+\bar V^{3/4},~~~
\bar z_m \leq\bar z \leq 1.
\label{3.31}
\eea
The electric field at the cathode, obtained by substituting
(\ref{3.30}) corresponding to $\bar z=0$ in (\ref{3.28}) reads
\bea
\bar E_0={4\sqrt{\alpha(2-f)}\over 3}.
\label{3.32n}
\eea
The unknowns $\bar z_m$ and $f$ in the expressions above are
determined by demanding that $\bar\phi$ in the two regions should
match at $\bar z=\bar z_m$. To that end, note that, by virtue of
(\ref{3.24}), the expressions (\ref{3.30}) and (\ref{3.31}), evaluated
at $\bar z=\bar z_m$, lead to the relations
\bea
\bar z_m&=&{1\over\sqrt{\alpha (2-f)}},
\label{3.32}\\
\bar z_m&=&1-{\bar V^{3/4}\over\sqrt{\alpha f}}.
\label{3.33}
\eea
These equations determine two unknowns, the position $\bar z_m$ of
the plane of reflection and the fraction $f$ of the injected current
that is transmitted. On eliminating $\bar z_m$ between these
equations, the expression determining $f$ reads
\bea
\sqrt{\alpha}={1\over\sqrt{2-f}}
+{\bar V^{3/4}\over\sqrt{f}}.
\label{3.34}
\eea
It is straightforward to verify that ${\rm d}f/{\rm d}\alpha<0$.
Hence, $f$ is a decreasing function of $\alpha$. Now, (\ref{3.34})
shows that the value $f=1$ is attained when
$\alpha=\alpha_{{\rm BF}}$ where
\bea
\alpha_{{\rm BF}}&=&\left(1+\bar V^{3/4}\right)^2.
\label{3.35}
\eea
According to classical theory, this is, therefore, the minimum value
of $\alpha$ for which reflecting solution exists. The subscript
${\rm BF}$ on $\alpha_{{\rm BF}}$ indicates that the steady state
bifurcates into two different types at $\alpha_{{\rm BF}}$: the steady
state corresponds to unidirectional flow of injected current $\alpha$
if $\alpha\le\alpha_{{\rm BF}}$ whereas it starts reflecting if
$\alpha>\alpha_{{\rm BF}}$. On comparing (\ref{3.35}) with (\ref{n1p})
it follows that $\alpha_{{\rm DR}}=\alpha_{{\rm SCL}}$ i.e. the
bifurcation of the steady state from unidirectional flow to partially
reflecting flow takes place at the same value of current above which
there are two solutions for the unidirectional flow, one stable and
another unstable. The electric field at the entrance of the diode for
$\alpha=\alpha_{{\rm BF}}$, obtained using (\ref{3.32n}), reads
\bea
\bar E_{0{\rm BF}}\equiv -\left(
{{\rm d}\bar\phi \over{\rm d}\bar z}\right)\Big|_{\bar z=0}
={4\over 3}(1+\bar V^{3/4}).
\label{efbf}
\eea
This is the same as the one in (\ref{efdr}) obtained using the
steady state solution for $\alpha=\alpha_{{\rm DR}}$.

It may be verified that the current $J_0$ corresponding to
$\alpha_{{\rm BF}}$ is given by (\ref{eq:bf}). Since the bifurcation
point marks the onset of reflections, which is one of the
characteristic properties of the Child-Langmuir current, the
current $\alpha_{{\rm BF}}$ at the bifurcation point is identified in
Ref.~\cite{liu} as the Child-Langmuir current. However, the numerical
solution of equations shows that reflection does not start at the
bifurcation point. It starts only when the injected current exceeds
its space charge limiting value $\alpha_{{\rm SCL}}$ which is higher
than its value at the bifurcation point except when $v_0=0$ in which
case, a comparison of (\ref{eq:scl}) and (\ref{eq:bf}) shows that
\bea
J_{{\rm BF}}=J_{{\rm SCL}},\qquad v_0=0.
\label{bn1}
\eea
On comparing with (\ref{eq:CL1}) it follows that these currents are
the same as the Child-Langmuir current $J_{{\rm CL}}$.

The behaviour of $\bar E_0$ for partially reflecting solution is
exhibited in Fig.1 by the branch of the curve above the bifurcation
point $B$. That branch is obtained by computing first the fraction
$f$ of the transmitted current by solving \ref{3.34} for a given
$\alpha$. The value of $f$ so obtained is substituted in \ref{3.32n}
to evaluate $\bar E_0$.

Before proceeding further, we digress to rewrite Eq.(\ref{3.33}) to
express the transmitted current $f\alpha$ in the form
\bea
f\alpha={\left(1+eV/\varepsilon\right)^{3/2}\over (1-\bar z_m)^2}.
\label{3.38}
\eea
This resembles the expression derived by Langmuir~\cite{langmuir}
for the thermionic current from a hot cathode emitting into a diode.
Langmuir's expression in question reduces exactly to (\ref{3.38}) if
the cathode temperature is zero whereas it gives the current
approximately if the temperature is non-zero but low. It should,
however, be emphasised that in the model under discussion, all the
electrons are injected in to the diode with the same velocity whereas,
in the paper of Langmuir, the electronic velocities at the entrance
are distributed according to the Boltzmann distribution. The position
$\bar z_m$ of the potential minimum in (\ref{3.38}) determining
$f\alpha$ itself depends on $f$. Hence, evaluation of (\ref{3.38})
requires a relation between $\bar z_m$ and $f$ which, in our case, is
contained in (\ref{3.32}).

The Eq.(\ref{3.34}) leads to the following useful expression for the
transmitted current,
\bea
\alpha_{{\rm TR}}\equiv f\alpha=\left[\bar V^{3/4}
+\sqrt{{f\over 2-f}}\right]^2,
\label{3.36}
\eea
for all $\alpha\ge \alpha_{{\rm BF}}$. In terms of the unnormalized
variables, this reads 
\bea
J_{\rm TR}&=&J_{\rm CL} \left[\left({mv_0^2\over 2eV}\right)^{3/4}
\sqrt{{f\over 2-f}} +
\left(1 + {mv_0^2\over 2eV} \right)^{3/4} \right]^2
\label{eq:tr}
\eea
for all $J_0\ge J_{{\rm BF}}$. Since $0\le f\le 1$, this expression,
in the limit of vanishing injection velocity
$(\varepsilon\rightarrow 0)$ yields
\bea
J_{\rm TR} = J_{\rm CL},\qquad v_0=0,\qquad J_0\ge J_{{\rm BF}},
\label{eq:tr2}
\eea
where $J_{\rm CL}$ is the Child-Langmuir current (\ref{eq:CL1}). This
is the current transmitted for any injected current above its space
charge limiting value. This value of current is the same as the value
of the space charge limiting current (\ref{eq:scl}) corresponding
to $v_0=0$. It therefore follows that, for $v_0=0$, the transmitted
current saturates at $J_{\rm CL}$ as soon as the injected current
attains its space charge limiting value.

The saturation current for non-zero injection velocities, obtained
from (\ref{3.36}) in the limit $f\rightarrow 0$, reads
\bea
\alpha_{{\rm TR}}\rightarrow \alpha_{{\rm GCL}}, ~~~~~
{\rm as}~~~f\rightarrow 0,
\label{3.37}
\eea
where $\alpha_{{\rm GCL}}$ is the saturation current given by
\bea
\alpha_{{\rm GCL}}=\bar V^{3/2}.
\label{3.37b}
\eea
We call the saturation current in the equation above as the
generalized Child-Langmuir current. Its unnormalized form is given
by (\ref{eq:gcl}).

The finite value of the transmitted current $f\alpha$ in the limit
$f\rightarrow 0$ implies that it is achieved when
$\alpha\rightarrow\infty$. Thus, the current transmitted across the
diode saturates at the value given by (\ref{3.37b})
when the current injected in it is increased indefinitely.

We assess the validity of the classical theory by comparing its
predictions with numerical results in the next section.

\section{Numerical Results}
In this section we present a comparison of the predictions of the
steady state theories and numerical computations carried out using
the sheet model \cite{birdsall}.
The comparison is carried by
computing the transmitted current $J_{{\rm TR}}$ as a function of
the injected current $J_0$ ($\alpha$ is related to $J_0$ by
Eq.~(\ref{alpha}) ).

The injected current $J_0$ in the sheet model 
corresponds to injecting $N$ sheets per unit time ($L/(2eV/m)^{1/2}$)
into the diode, each having a surface charge density $\sigma = J_0/N$.
By counting the sheets starting from the one closest to the collecting
plate, it can be shown that when there are $M$ sheets in the drift
space, the electric field on the $k^{{\rm th}}$ sheet at the position
$z_k$ from the entrance plate is given by

\bea
E(z_k)&=&-4\pi\sigma\left[k-{1\over L}\sum_{j=1}^{M}z_j
-{1\over 2}\right] + {V \over L}.
\eea

\noindent
The (unnormalized) potential at any point $z$ between the
$(k)^{th}$ and $(k+1)^{th}$ sheet is similarly given by

\bea
\phi(z) & = & {V z \over L} - {4\pi J_0 \over N} \left[
z\left(k L - \sum_{i=1}^{M} z_i \right) + L \sum_{i=k+1}^{M} z_i \right]
\eea

\noindent
where the sheet number is counted from the anode (right) plate.
The force equation for the sheet is thus

\bea
m {{\rm d}^2 z_k\over{\rm d} t^2} = e E(z_k)
\eea

\noindent
The numerical procedure consists of solving the coupled set of
equations for $ z_i(t)$, ($i=1,2,\ldots$) as a
function of time by choosing the time-step $\Delta t$ of
integration and the rate of injection $N$ of the
sheets in such a way that desired accuracy is achieved.
Typically, $\Delta t = 8.532 \times 10^{-13}$s, $N = 1000$  for
$J_0 = 20000 {\rm A/m^2} $

The numerical computations confirm the theoretical prediction that,
for $\alpha\le\alpha_{{\rm SCL}}$, the system approaches a steady
state asymptotically as $t\rightarrow\infty$. However, contrary to the
premise of the classical theory, the asymptotic behaviour of the
system in time is found to be oscillatory if
$\alpha>\alpha_{{\rm SCL}}$. Hence, for $\alpha>\alpha_{{\rm SCL}}$,
the predictions of the classical theory are compared with the results
of numerical computations averaged over several cycles of
oscillations after the initial transients have settled. 
See 
Ref.~\cite{dubinov} for a plot of oscillating transmitted current as a
function of time for injected current in excess of the SCL current in
a short-circuited diode.

The Fig.2 is a plot of theoretical predictions and numerically
computed values of the transmitted current $J_{{\rm TR}}$ as a
function of the injected current $J_0$ for $\varepsilon=0$ and for
$\varepsilon/eV=0.2$ with $L=.008$m and $V=250$V. The bold straight
line represents $J_{{\rm TR}}=J_0$ which is the prediction of the
steady state theory for currents below the space charge limiting
value. The square dots on that line are the results of numerical
computations indicating agreement between analytical and numerical
solutions for currents below the space charge limiting value.

The numerical results of computation for currents exceeding the space
charge, carried in the manner outlined above, are represented by the
square dots. The dashed-dotted straight line (a) is drawn to join the
dots corresponding to $\epsilon=0$ whereas the dashed-dotted curve (b)
for $\varepsilon/eV=0.2$ joins the dots corresponding to
$\varepsilon=0.2$. Recall that the classical theory predicts that if
$\varepsilon=0$ then $J_{{\rm TR}}=J_{{\rm CL}}$ for all $J_0$. For
the values of the parameters used in the figure, the value of the
Child-Langmuir current is $J_{{\rm CL}}\simeq 143.9$ which
practically coincides with the corresponding numerical results
represented by the straight line (a). Hence, the theoretical
predictions are in agreement with numerical results for any value of
the injected current if $\varepsilon=0$.

The theoretical plot for $\varepsilon/eV=0.2$ is represented by the
dashed curve marked ``classical theory". It is obtained by solving
(\ref{3.34}) for $f$ for a given value of $\alpha$ and using the
value of $f$ so obtained to evaluate the transmitted current
$J_{{\rm TR}}=f\alpha$. On comparing the classical theory plot with the
corresponding results of numerical computations represented by the
curve (b), it is seen that the agreement between the two is not close.
For, in disagreement with the monotonic behaviour predicted by the
classical theory, the numerical plot exhibits a maxima in
$\alpha_{{\rm TR}}$ as a function of $\alpha$. 
The reason behind 
the appearance of the maxima is left as an open question. However, the
asymptotic limit of the numerical plot of $\alpha_{{\rm TR}}$ appears
to be in agreement with the prediction (\ref{3.37}) of the classical
theory. In order to confirm the apparent agreement we examine the
asymptotic behaviour of the curve
\bea
J_{{\rm TR}}=a+bJ^{-\beta}_{{\rm 0}}
\label{fit}
\eea
fitting best the numerical results. The Fig.3 shows some values
of $J_{{\rm TR}}$ computed numerically along with the curve
(\ref{fit}) fitting best the numerical results as a function of
$J_{0}$. The best fit is obtained for $a = 193.58$, $b = 2318.83$ and
$\beta = 0.40$. Hence, (\ref{fit}) approaches asymptotically the value
$J_{{\rm TR}} = 193.58$ which agrees very well with the value
$J_{{\rm TR}} = 189.37$ obtained using the theoretical expression
(\ref{3.37b}). We have observed similar agreement between theoretical
prediction and the numerical computations for several other values
of the injected velocity.

Finally, we study the behaviour of the (time-averaged) position 
of the virtual cathode as a function of the
injected current $J_0$ 
for $\varepsilon = 0.2$. Fig.~4 shows a plot of the time-averaged
normalized position of the potential minimum ($\bar{z}_m$) as
a function of $J_0$. As in the case of short-circuited
diode, $\bar{z}_m$ decreases with $J_0$. Fig.~4 also shows the
prediction of the classical theory (bold line) and the
best fit of the function $a_1 + b_1 J^{-\delta}_0$ (dashed line)
for $J_0$ in the interval [1000,20000].
The value of the parameters are :
$a_1 = -0.007$, $b_1 = 1.384$ and $\delta = 0.335$. Clearly, the position
of the virtual cathode moves to the injection plane as $J_0 \rightarrow
 \infty$.

\section{Conclusions} 

In this paper, we have sought to generalize the Child-Langmuir
law for non-zero injection velocities based on the saturation
property of the time-averaged transmitted current. The reason for
preferring the saturation property for the said generalization is
its importance in numerical simulations. For, as stated in the
Introduction, the emission characteristics of cathode and the
details of the emission mechanism do not play any role in
determining the transmitted current if that current is
independent of (or weakly dependent on) the injected current. This
is the case when the current emitted from the cathode surface ($J_0$) 
is more than what gives rise to the saturation value of the transmitted
current. This condition simplifies considerably the task of numerical
simulation.

Thus, if we characterize the Child-Langmuir current as the saturation
current then $\alpha_{{\rm GCL}}$ given in Eq.~(\ref{3.37b}) is the
Child-Langmuir current generalized to non-zero velocities of injection.
That expression for the saturation current is derived by assuming a
steady flow of current under the condition that a part of the current
is reflected from the virtual cathode. Eq.~(\ref{3.37b}) is
found to be in fairly good agreement with numerical computations. In
contrast to the case of zero velocity of injection for which
saturation is attained at the space charge limiting value of the
injected current, the saturation in the case of non-zero velocities is
attained only asymptotically. Hence, though the details of the
emission characteristic of cathode may be ignored safely in numerical
simulations for any injected current above its space charge limiting
value in case $v_0=0$, the injected current for $v_0\ne 0$ needs to be
very much above its space charge limiting value for cathode
characteristics to become unimportant.

\appendix\section{Steady state in Lagrangian formalism}
In this Appendix we present steady state solution of the
Eq.(\ref{2.9}) describing the flow of current in Lagrangian formalism.

Integration of (\ref{2.9}) yields
\bea
{\partial v(t,t_0)\over\partial t}=-{e\over m}\left[E_0(t)
+4\pi\int_{t_0}^{t}~{\rm d}\tau J_0(\tau)\right].
\label{2.11}
\eea
This equation is arrived at by noting that the electron at time
$t=t_0$ is at $z=0$. Hence, the initial condition is the same as the
condition at $z=0$ so that, by virtue of the last equation in
(\ref{2.1}), $\partial v(t,t_0)/\partial t=-eE_0/m$ at
$t=t_0$. The equation (\ref{2.11}) may be integrated further to obtain
expressions for $v(t,t_0)$ and $z(t,t_0)$. Like (\ref{2.11}), those
equations involve two unknown functions: the field and the current at
all times at the entrance plate. The current at the entrance plate
consists of the current being constantly pumped in and the current
returning, if any, after reflection from some position between the
plates. Our interest here is in the steady state without returning
currents. The current at the entrance plate in that case is the same
as that being pumped in i.e.
\bea
J_0(t)=J_0=\rho_0v_0=-|\rho_0|v_0.
\label{2.12}
\eea
We normalize the variables as in (\ref{2.13}) and integrate
(\ref{2.11}) in the absence of returning currents
to get
\bea
\bar v(\bar t,\bar t_0)&=&1-{1\over 2}\int_{\bar t_0}^{\bar t}
~{\rm d}\tau~ \bar E_0(\tau){\rm d}\tau+{\alpha\over 9}
(\bar t-\bar t_0)^2,
\label{2.14}\\
\bar z(\bar t,\bar t_0)&=&(\bar t-\bar t_0)
-{1\over 2}\int_{\bar t_0}^{\bar t}
~{\rm d}\tau~(\bar t-\tau)\bar E_0(\tau)\nonumber\\
&&+{\alpha\over 27}(\bar t-\bar t_0)^3.
\label{2.15}
\eea
The unknown function $\bar E_0(t)$ in these equations is found by
imposing the boundary conditions (\ref{2.6}). That task is simplified
in the steady state because the field $E_0(t)$ at the entrance plate
is then independent of time reducing (\ref{2.14}) and (\ref{2.15}) to
the forms which are functions only of the time difference $t-t_0$:
\bea
\bar v(\bar t,\bar t_0)&=&1-{1\over 2}(\bar t-\bar t_0)
\bar E_0+{\alpha\over 9}(\bar t-\bar t_0)^2,
\label{3.6}\\
\bar z(\bar t,\bar t_0)&=&(\bar t-\bar t_0)
-{1\over 4}(\bar t-t_0)^2\bar E_0
+{\alpha\over 27}(\bar t-\bar t_0)^3.
\label{3.7}
\eea
Note that these equations already incorporate the boundary condition
at $z=0$. The boundary condition (\ref{2.6}) at $z=L$, restated
in terms of the velocity by invoking the conservation law (\ref{3.3}),
leads to the expression (\ref{3.8}) for velocity at the collector
plate in terms of its potential.

Now, if $\bar T(1)$ is the time that an electron takes to reach the
collector plate after its entry in to the diode then the Equations
(\ref{3.6}) and (\ref{3.7}) yield
\bea
\bar v_1&=&1-{1\over 2}\bar T(1) \bar E_0+{\alpha\over 9}\bar T^2(1),
\label{3.9}\\
1&=&\bar T(1)-{1\over 4}\bar T^2(1)\bar E_0
+{\alpha\over 27}\bar T^3(1).
\label{3.10}
\eea
The two equations above determine the unknowns $\bar E_0$ and
$\bar T(1)$. On eliminating $\bar E_0$ between these equation, the
transit time $\bar T(1)$ is found to obey the cubic
\bea
\bar T^3(1)-{27\over\alpha}(1+\bar v_1)\bar T(1)
+{54\over\alpha}=0.
\label{3.11}
\eea
Of course, only the real positive roots of this equation are
physically acceptable. The electric field at the entrance plate,
obtained by substituting an acceptable solution of (\ref{3.11}) in
(\ref{3.9}) is
\bea
\bar E_0={2\over\bar T(1)}\left[{\alpha\over 9}\bar T^2(1)+1
-\bar v_1\right].
\label{3.11b}
\eea
Now, the nature of the roots is determined by the sign of the
discriminant
\bea
\Delta={1\over 4}q^2+{1\over 27}p^3,
\label{3.12}
\eea
where
\bea
p=-{27\over\alpha}(1+\bar v_1),\qquad q={54\over\alpha}
\label{3.13}
\eea
so that
\bea
\Delta=\left({27\over\alpha}\right)^2
\left(1-{1\over\alpha}(1+\bar v_1)^3\right).
\label{3.15}
\eea
If $\Delta\le 0$ then the three roots are real whereas one root is
real and the other two complex conjugate of each other if $\Delta>0$.

Note that $\Delta\le 0$ and hence the three roots are real if
$\alpha\le (1+\bar v_1)^3$. The roots in this case are
\bea
\bar T_1(1)&=&-2A^{1/3}\cos\left({\theta\over 3}\right),\nonumber\\
\bar T_2(1)&=&2A^{1/3}\cos\left({\theta+\pi\over 3}\right)\nonumber\\
\bar T_3(1)&=&2A^{1/3}\cos\left({\pi-\theta\over 3}\right),
\label{3.16}
\eea
where
\bea
A&=&27\left[{1\over\alpha}(1+\bar v_1)\right]^{3/2},\nonumber\\
{\rm tan}(\theta)&=&\sqrt{\left({1\over\alpha}
\left(1+\bar v_1\right)^3-1\right)},
\quad 0\le\theta\le{\pi\over 2}.
\label{3.17}
\eea
The root $\bar T_1(1)$, being negative, is physically unacceptable.
Hence, $\bar T_2(1)$ and $\bar T_3(1)$ are the admissible times of
transit. The expression for the electric field (\ref{3.11b}) at the
entrance of the cathode due to the positive roots assumes the form
\bea
\bar E_0&=&{2\sqrt{\alpha}\over 3(1+\sqrt{\bar v_1})}
\left[\cos\left({\pi\pm\theta\over 3}\right)\right]^{-1}\nonumber\\
&&\times\left[4(1+\bar v_1)
\cos^2\left({\pi\pm\theta\over 3}\right)+1-\bar v_1\right].
\label{3.17b}
\eea
We now examine the question of admissibility of $\bar T_2(1)$ and
$\bar T_3(1)$ based on other physics considerations. We will see that
the slower of the two transit times, $\bar T_3(1)$, corresponds to
negative velocity in some region if
$0\le\alpha\le(1+\bar v^{3/2}_1)^2$ and to positive velocity
all through the diode but an unstable steady state if
$(1+\bar v^{3/2}_1)^2<\alpha\le (1+\bar v_1)^3$.

To that end, let $\bar v_m$ be the minimum velocity at some place in
the diode and let $\bar T_m$ be the time that an electron takes to
reach that position so that ${\rm d}\bar v/{\rm d}\bar T=0$ at
$\bar T=\bar T_m$. On solving (\ref{2.14}) (with $\bar t-\bar t_0
=\bar T$) for ${\rm d}\bar v/{\rm d}\bar T=0$, it follows that
\bea
\bar T_m&=&{9\over 4\alpha}\bar E_0,
\label{3.18}\\
\bar v_m&=&1-{9\over 16\alpha}\bar E^2_0.
\label{3.19}
\eea
This equation shows that the minimum velocity is negative if either
$\bar E_0<-4\sqrt{\alpha}/3$ or $\bar E_0>4\sqrt{\alpha}/3$. However,
(\ref{3.6}) shows that if $\bar E_0\le 0$ then $\bar v\ge 0$. Hence,
the condition for the velocity to become negative in some region in
the diode is $\bar E_0>4\sqrt{\alpha}/3$ which, in conjunction with
(\ref{3.11b}),
reads
\bea
\left({\sqrt{\alpha}\over 3}\bar T(1)-1-\sqrt{\bar v_1}\right)
\left({\sqrt{\alpha}\over 3}\bar T(1)-1+\sqrt{\bar v_1}\right)>0.
\label{3.19b}
\eea
The positivity of $\bar T(1)$ implies that (\ref{3.19b}) can be
satisfied if
\bea
\bar T(1)>{3\over\sqrt{\alpha}}(1+\sqrt{\bar v_1}).
\label{3.20b}
\eea
This condition, for $\bar T(1)=\bar T_2(1),\bar T_3(1)$
given in (\ref{3.16})), assumes the form
\bea
\cos\left({\pi\pm\theta\over 3}\right)>
{1\over 2}\left(1+\sqrt{\bar v_1}\right)
\left(1+\bar v_1\right)^{-1/2}.
\label{3.21}
\eea
It is easy to check that this implies $\cos((\pi\pm\theta)/3))>1/2$
which, in turn, implies $(\pi\pm\theta)/3<\pi/3$. Since $\theta$ is
positive, the last inequality can be satisfied only by $\pi-\theta$.
Thus, the minimum velocity can become negative only for $\bar T_3(1)$.
Hence, $\bar T_2(1)$ is the only acceptable root for those values of
$\alpha$ for which $\bar T_3(1)$ leads to negative velocities. Those
values of $\alpha$ may be identified by noting that the condition
(\ref{3.21}) for $\bar T_3(1)$ implies
\bea
\pi-\theta<3\cos^{-1}\left[{1\over 2}\left(1+\sqrt{\bar v_1}
\right)\left(1+\bar v_1\right)^{-1/2}\right].
\label{3.21b}
\eea
Using the trigonometric relation $\cos(3x)=4\cos^3(x)-3\cos(x)$,
and the defining relation (\ref{3.17}) of $\theta$, it is
straightforward to show that (\ref{3.21b}) holds if
\bea
\alpha\le (1+\bar v_1^{3/2})^2\equiv\alpha_{{\rm DR}},
\label{3.23}
\eea
where $\alpha_{{\rm DR}}$ is given by (\ref{n1p}). In other words, if
$\alpha$ satisfies the condition (\ref{3.23}) then the root
$\bar T_3(1)$ satisfies the condition (\ref{3.20b}) required for the
velocity to become negative in some region in the diode leaving
$\bar T_2(1)$ as the only acceptable root in the range
$0\le\alpha\le\alpha_{{\rm DR}}$ whereas both, $\bar T_2(1)$ and
$\bar T_3(1)$, are acceptable if $\alpha_{{\rm DR}}\le\alpha<
(1+\bar v_1)^3$. The $\alpha_{{\rm DR}}$ denotes the value of the
current above which there are two roots. However, the linear stability
analysis shows that the root $\bar T_3(1)$ is unstable. Hence, only
$\bar T_2(1)$ corresponds to a stable steady state in the range
$0\le\alpha\le(1+\bar v_1)^3$.

The discriminant $\Delta>0$ if $\alpha>(1+\bar v_1)^3$. The
cubic (\ref{3.11}) determining the transit time in this case has one
real and a pair of complex conjugate roots. Since the transit time
is a real quantity, the real root is the only one which needs to be
examined further for its acceptability. To that end, we note that
the product of the roots of (\ref{3.11}), given by $-54/\alpha$,
is negative. Hence, the real root in case of $\Delta>0$ must be
negative. Since the time of transit is a positive number, the negative
root is physically unacceptable. The cubic (\ref{3.11}), therefore,
does not admit physically acceptable solution i.e. the system does not
approach a time-independent steady state having unidirectional flow of
current if
$\alpha>\alpha_{{\rm SCL}}$ where
\bea
\alpha_{{\rm SCL}}=(1+\bar v_1)^3
\label{scl}
\eea
is the space charge limiting current: The space charge limiting
current is the value of the current above which the flow of current is
no longer unidirectional and time-independent. It is also the maximum
current that can be transmitted across the diode even when the
injected current is increased indefinitely.

The solution of (\ref{3.11}) for $\alpha=\alpha_{{\rm SCL}}$ gives the
transit time as
\bea
T_{{\rm SCL}}(1)={3\over 1+\sqrt{\bar V}},
\label{3.15c}
\eea
which, on substitution in (\ref{3.11b}) yields
\bea
\bar E_{{\rm SCL}0}={4\over 3}(1+\sqrt{\bar V}).
\label{3.15d}
\eea
This is the electric field at the entrance of the diode when the
value of the injected current $\alpha=\alpha_{{\rm SCL}}$.


\newcommand{\PR}[1]{{Phys.\ Rep.}\/ {\bf #1}}
\newcommand{\PRL}[1]{{Phys.\ Rev.\ Lett.}\/ {\bf #1}}
\newcommand{\PRA}[1]{{Phys.\ Rev.\ A}\/ {\bf #1}}
\newcommand{\PRB}[1]{{Phys.\ Rev.\ B}\/ {\bf #1}}
\newcommand{\PRD}[1]{{Phys.\ Rev.\ D}\/ {\bf #1}}
\newcommand{\PRE}[1]{{Phys.\ Rev.\ E}\/ {\bf #1}}
\newcommand{\JPA}[1]{{J.\ Phys.\ A}\/ {\bf #1}}
\newcommand{\JPB}[1]{{J.\ Phys.\ B}\/ {\bf #1}}
\newcommand{\JCP}[1]{{J.\ Chem.\ Phys.}\/ {\bf #1}}
\newcommand{\JPC}[1]{{J.\ Phys.\ Chem.}\/ {\bf #1}}
\newcommand{\JMP}[1]{{J.\ Math.\ Phys.}\/ {\bf #1}}
\newcommand{\JSP}[1]{{J.\ Stat.\ Phys.}\/ {\bf #1}}
\newcommand{\AP}[1]{{Ann.\ Phys.}\/ {\bf #1}}
\newcommand{\PLB}[1]{{Phys.\ Lett.\ B}\/ {\bf #1}}
\newcommand{\PLA}[1]{{Phys.\ Lett.\ A}\/ {\bf #1}}
\newcommand{\PD}[1]{{Physica D}\/ {\bf #1}}
\newcommand{\NPB}[1]{{Nucl.\ Phys.\ B}\/ {\bf #1}}
\newcommand{\INCB}[1]{{Il Nuov.\ Cim.\ B}\/ {\bf #1}}
\newcommand{\JETP}[1]{{Sov.\ Phys.\ JETP}\/ {\bf #1}}
\newcommand{\JETPL}[1]{{JETP Lett.\ }\/ {\bf #1}}
\newcommand{\RMS}[1]{{Russ.\ Math.\ Surv.}\/ {\bf #1}}
\newcommand{\USSR}[1]{{Math.\ USSR.\ Sb.}\/ {\bf #1}}
\newcommand{\PST}[1]{{Phys.\ Scripta T}\/ {\bf #1}}
\newcommand{\CM}[1]{{Cont.\ Math.}\/ {\bf #1}}
\newcommand{\JMPA}[1]{{J.\ Math.\ Pure Appl.}\/ {\bf #1}}
\newcommand{\CMP}[1]{{Comm.\ Math.\ Phys.}\/ {\bf #1}}
\newcommand{\PRS}[1]{{Proc.\ R.\ Soc. Lond.\ A}\/ {\bf #1}}
\newcommand{\PP}[1]{{Phys.\ Plasmas\ }\/ {\bf #1}}


\vskip -1.0 in

\begin{center}
\begin{figure}[tbp]
\hspace*{0.1cm}\epsfig{figure=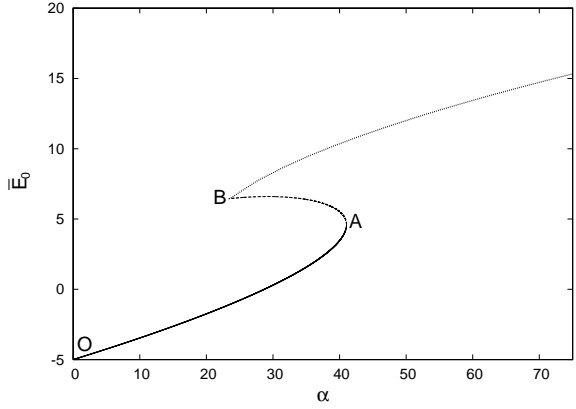,width=5.5cm,angle=270}
\vskip 0.25 in
\caption[ty] {The electric field at the cathode $\bar E_0$ plotted
as a function of the current $\alpha$ for $\varepsilon/eV=0.2$.
The branches OA (stable) and AB (unstable) correspond respectively to
$\bar T_2(1)$ and $\bar T_3(1)$ given by (\ref{3.16}). The uppermost
branch of the curve for $\alpha>\alpha_{{\rm DR}}
(=\alpha_{{\rm BF}})$ (point B) is a result of the classical theory.}
\label{fig:1}
\end{figure}
\end{center}


\begin{center}
\begin{figure}[tbp]
\hspace*{0.1cm}\epsfig{figure=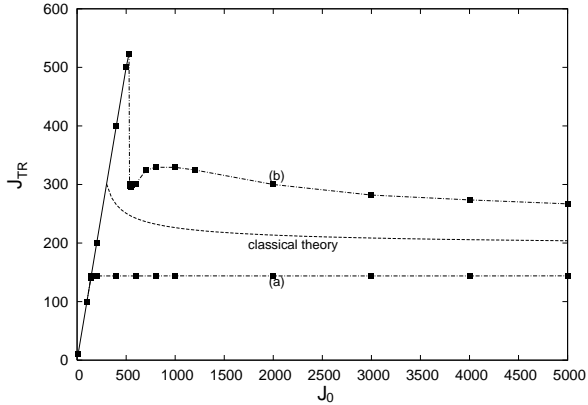,width=5.5cm,angle=270}
\vskip 0.25 in
\caption[ty]{The transmitted current density, $J_{{\rm TR}}$, plotted
against the injected current density, $J_{0}$, for two different value
of $\varepsilon/eV$ with $L=0.008$m $V=250$V. The current density is
in units of ${\rm Amp/m}^2$. The square dots are the results of the
numerical computation. The bold line is $J_{{\rm TR}}=J_{0}$. The
straight line  (a) is drawn to join the results of numerical
computations for $\varepsilon=0$ while the curve (b) joins the
numerical results for $\varepsilon/eV = 0.2$. The dotted line marked
``classical theory" is the prediction of classical theory for
$\varepsilon/eV = 0.2$.}
\label{fig:2}
\end{figure}
\end{center}


\begin{center}
\begin{figure}[tbp]
\hspace*{0.1cm}\epsfig{figure=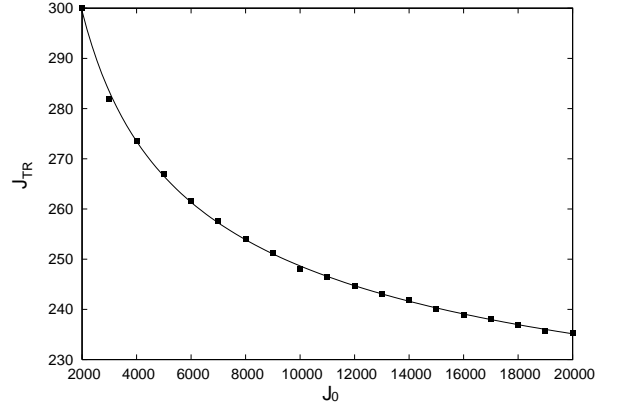,width=5.5cm,angle=270}
\vskip 0.25 in
\caption[ty]{A fit of the transmitted current density using
$J_{{\rm TR}}=a+bJ^{-\beta}_{{\rm 0}}$ as a function of the injected
current density $J_0$. The current density is units of
${\rm Amp/m}^2$. The solid line is the best fit while the squares
mark the numerically computed values of the transmitted current using
the sheet model.}
\label{fig:3}
\end{figure}
\end{center}


\begin{center}
\begin{figure}[tbp]
\hspace*{0.1cm}\epsfig{figure=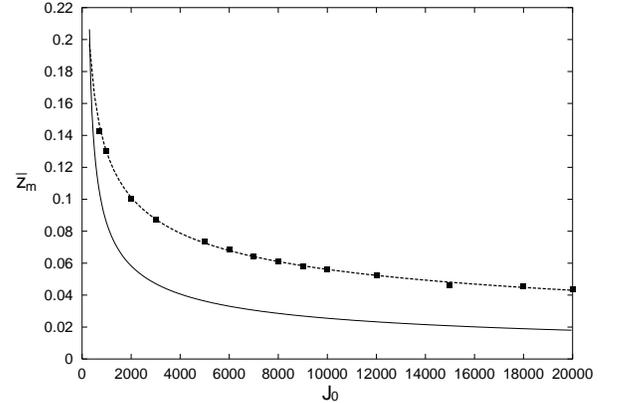,width=5.5cm,angle=270}
\vskip 0.25 in
\caption[ty]{A plot of the time-averaged
normalized position of the virtual cathode (solid squares) as
a function of $J_0$.  Also shown are the prediction of the
classical theory (bold line) and the best fit using the
function $a_1+b_1J^{-\delta}_0$ (dashed line).
}
\label{fig:4}
\end{figure}
\end{center}



\end{document}